\documentclass[11pt]{amsart}

\usepackage[
hypertexnames=false, colorlinks, citecolor=red, linkcolor=red]
{hyperref} 
\hypersetup{bookmarksdepth=3}

\usepackage{amsmath, amsthm, amssymb}
\usepackage{graphicx}
\usepackage{epstopdf}
\usepackage{amssymb}
\usepackage{url}
\usepackage{mathrsfs}

	\normalbaselines						  %

\newcommand{\Z}{{\mathbb  Z}}

\newcommand{\N}{{\mathbb  N}}

\newcommand{\C}{{\mathbb  C}}

\newcommand{\fdot}{\,\cdot\,}

\makeatletter
\def\Ddots{\mathinner{\mkern1mu\raise\p@
\vbox{\kern7\p@\hbox{.}}\mkern2mu
\raise4\p@\hbox{.}\mkern2mu\raise7\p@\hbox{.}\mkern1mu}}
\makeatother

\newcommand{\cH}{\mathcal{H}}

\newcommand{\p}{\mathbb{P}}

\DeclareMathOperator{\spa}{span}

\DeclareMathOperator{\clos}{clos}

\DeclareMathOperator{\dist}{dist}

\newcommand{\ci}[1]{_{ {}_{\scriptstyle #1}}}
\newcommand{\ti}[1]{_{\scriptstyle \text{\rm #1}}}

\catcode`@=11
\chardef\mathlig@atcode\count255

\def\actively#1#2{\begingroup\uccode`\~=`#2\relax\uppercase{\endgroup#1~}}
\def\mathlig@gobble{\afterassignment\mathlig@next@cmd\let\mathlig@next= }
\def\mathlig@delim{\mathlig@delim}
\def\mathlig@defcs#1{\expandafter\def\csname#1\endcsname}
\def\mathlig@let@cs#1#2{\expandafter\let\expandafter#1\csname#2\endcsname}
\def\mathlig@appendcs#1#2{\expandafter\edef\csname#1\endcsname{\csname#1\endcsname#2}}

\def\mathlig#1#2{\mathlig@checklig#1\mathlig@end\mathlig@defcs{mathlig@back@#1}{#2}\ignorespaces}

\def\mathlig@checklig#1#2\mathlig@end{%
 \expandafter\ifx\csname mathlig@forw@#1\endcsname\relax
 \expandafter\mathchardef\csname mathlig@back@#1\endcsname=\mathcode`#1%
 \mathcode`#1"8000\actively\def#1{\csname mathlig@look@#1\endcsname}%
 \mathlig@dolig#1\mathlig@delim
\fi
\mathlig@checksuffix#1#2\mathlig@end
}

\def\mathlig@checksuffix#1#2\mathlig@end{%
\ifx\mathlig@delim#2\mathlig@delim\relax\else\mathlig@checksuffix@{#1}#2\mathlig@end\fi
}
\def\mathlig@checksuffix@#1#2#3\mathlig@end{%
\expandafter\ifx\csname mathlig@forw@#1#2\endcsname\relax\mathlig@dosuffix{#1}{#2}\fi
\mathlig@checksuffix{#1#2}#3\mathlig@end
}

\def\mathlig@dosuffix#1#2{%
\mathlig@appendcs{mathlig@toks@#1}{#2}%
\mathlig@dolig{#1}{#2}\mathlig@delim
}

\def\mathlig@dolig#1#2\mathlig@delim{%
 \mathlig@defcs{mathlig@look@#1#2}{%
 \mathlig@let@cs\mathlig@next{mathlig@forw@#1#2}\futurelet\mathlig@next@tok\mathlig@next}%
 \mathlig@defcs{mathlig@forw@#1#2}{%
  \mathlig@let@cs\mathlig@next{mathlig@back@#1#2}%
  \mathlig@let@cs\checker{mathlig@chck@#1#2}%
  \mathlig@let@cs\mathligtoks{mathlig@toks@#1#2}%
  \expandafter\ifx\expandafter\mathlig@delim\mathligtoks\mathlig@delim\relax\else
  \expandafter\checker\mathligtoks\mathlig@delim\fi
  \mathlig@next
 }%
 \mathlig@defcs{mathlig@toks@#1#2}{}%
 \mathlig@defcs{mathlig@chck@#1#2}##1##2\mathlig@delim{%
  \ifx\mathlig@next@tok##1%
   \mathlig@let@cs\mathlig@next@cmd{mathlig@look@#1#2##1}\let\mathlig@next\mathlig@gobble
  \fi
  \ifx\mathlig@delim##2\mathlig@delim\relax\else
   \csname mathlig@chck@#1#2\endcsname##2\mathlig@delim
  \fi
 }
  \ifx\mathlig@delim#2\mathlig@delim\else
  \mathlig@defcs{mathlig@back@#1#2}{\csname mathlig@back@#1\endcsname #2}%
 \fi
}%

\catcode`@\mathlig@atcode

\mathchardef\ordinarycolon\mathcode`\:
\def\vcentcolon{\mathrel{\mathop\ordinarycolon}}
\mathlig{:=}{\vcentcolon=}
\mathlig{::=}{\vcentcolon\vcentcolon=}

\newtheorem{theo}{Theorem}[section]

\newtheorem{prop}[theo]{Proposition}
\newtheorem{mr}[theo]{Main Result}

\newtheorem{criterion}[theo]{Numerical Criterion}

\newtheorem{rem}[theo]{Remark}

\begin{document}
\title[Delocalization at weak disorder]{Delocalization for the 3--D discrete random Schr\"odinger operator at weak disorder}
\author{Westin King}
\author{Robert C.~Kirby}
\author{Constanze Liaw}
\address{Department of Mathematics, Baylor University, One Bear Place $\#$97328, Waco, TX  76798-7328, USA}
\email{Westin$\underline{\,\,\,\,}$King@baylor.edu, Robert$\underline{\,\,\,\,}$Kirby@baylor.edu, Constanze$\underline{\,\,\,\,}$Liaw@baylor.edu}
\thanks{The work of Liaw and King was supported by DMS-1261687.}


\begin{abstract}
We apply a recently developed approach \cite{2D} to study the existence of extended states for the three dimensional discrete random Schr\"odinger operator at small disorder. The conclusion of delocalization at small disorder agrees with other numerical and experimental observations (see e.g.~\cite{PhysicsToday2009}). Further the work furnishes a verification of the numerical approach and its implementation.\\
Not being based on scaling theory, this method eliminates problems due to boundary conditions, common to previous numerical methods in the field. At the same time, as with any numerical experiment, one cannot exclude finite-size effects with complete certainty. Our work can be thought of as a new and quite different use of Lanczos' algorithm; a posteriori tests to show that the orthogonality loss is very small.\\
We numerically track the ``bulk distribution" (here: the distribution of where we most likely find an electron) of a wave packet initially located at the origin, after iterative application of the discrete random Schr\"odinger operator.
\end{abstract}

\maketitle

\section{Introduction}

Consider the discrete three dimensional Schr\"odinger operator, given by:
\begin{align}\label{d-RandSchr}
-\bigtriangleup f (x) = - \sum\limits_{|i|=1} (f(x+i)-f(x)),
\end{align}
when $i$ is of the form $(i_1,i_2,i_3)^T, i_k\in\Z$, and consider an element $\delta_i(x)$ of $l^2(\Z^3)$ given by
\begin{align*}
\delta_i(x)=
\left\{\begin{array}{ll}1&x=i\in\Z^3,\\ 0&\text{else.}\end{array}\right.
\end{align*}

Let the random variables $\omega_i$ be i.i.d.~with uniform distribution in $[-c/2,c/2]$, i.e.~according to the probability distribution $\mathbb{P} = c^{-1} \Pi_i  \chi\ci{[-c/2,c/2]} dx$.

The 3--D random discrete Schr\"odinger operator, formally given by
$$
H_\omega = - \bigtriangleup + \sum_{i\in \Z^3} \omega_i <\fdot, \delta_i> \delta_i
\text{ on }l^2(\Z^3),
$$
is the main object of study.

This operator has been studied extensively, see e.g.~\cite{Kirsh, SIMREV} and the references therein. The first part of the operator $- \bigtriangleup$ describes the movement of an electron inside a crystal with atoms located at all integer lattice points $\Z^3$. The perturbation $\sum_{i\in \Z^3} \omega_i <\fdot, \delta_i> \delta_i$ can be interpreted as having the atoms randomly displaced around the lattice points. It is important to notice that the perturbation is almost surely non-compact, so that classical perturbation theory (e.g.~Kato--Rosenblum Theorem, which states the invariance of the absolutely continuous spectrum under compact perturbations) cannot be applied almost surely. It is known that the absolutely continuous spectrum is deterministic, i.e.~it occurs with probability one or zero, see e.g.~\cite{Liaw2010}. 
Localization in the sense of exponentially decaying eigenfunctions was proved analytically for disorders $c$ \emph{above} some threshold $C_0$ (see e.g.~\cite{AizMol1993}, \cite{FS}, and \cite{SIMREV}). Currently, the smallest threshold in 3 dimensions is $C_0 = 100.6$ (see Table 1 in \cite{Schenker2013}).

Diffusion is expected but not proved for small disorder $c>0$. We numerically determine a regime of disorders for which the three dimensional discrete random Schr\"odinger operator does not exhibit localization. Our calculations are based on the Lanczos
algorithm~\cite{lanczos1950iteration} for determining orthogonal bases
for Krylov spaces~\cite{trefbau}.  Although we are not
the first to use this method (see e.g.~\cite{PhysicsToday2009, SteinKrey1979} and the references therein), our application of it is quite different.
In particular, our method is not based on scaling theory (for further discussion see \cite{2D}). In~\cite{SIAM2008}, the Lanczos algorithm is employed to compute a set of
eigenvalues and eigenvectors.
However,
we test for localization without computing eigenvalues or eigenvectors,
but only compute the distance between $\delta_{111}$ and the orbit of
$\delta_{000}$. The orbit is the span of $\left\{  H_\omega ^k
\delta_{000}: k\in\N\cup\{0\} \right\}$, which is exactly a Krylov
subspace.  At each step of the Lanczos iteration,
we use the orthogonality of the generated vectors to update the
distance of interest.  In this way, we maintain the low memory cost of
a three-term recurrence, bypassing the need to store any eigenvectors
at all.  In addition to this, we have performed some \emph{a
posteriori} tests of the Lanczos algorithm on smaller cases to measure
the degree to which orthogonality may be lost.

Besides computational advantages, our approach also offers a different
mathematical perspective.  By utilizing eigenvectors, it is (tacitly)
assumed that all spectral points are in fact eigenvalues, while our
approach merely generates an orbit without attempting to rule out other
kinds of spectral points.

While the contributions of this paper are numeric, the method (see \cite{2D}) provides an explicit analytic expression, which may yield a proof of the following numerically supported Main Result.

\begin{mr}\label{t-mr}
For disorder $c \lesssim 3.5$, numerical experiments indicate that the three dimensional discrete random Schr\"odinger operator does not exhibit Anderson localization with positive probability, in the sense that it has non-zero absolutely continuous spectrum with probability 1.
(In particular, we do not have what is usually referred to as ``strong dynamical localization" implying delocalization in most or even all of the other senses, see \cite{hundertmark}.)
\end{mr}

The key analytical tool to our method is stated in Proposition \ref{c-tool} below.
Section \ref{s-setup} is devoted to a description of the numerical experiment. The numerical testing criterion we applied is given by Numerical Criterion \ref{nc} below.
Our numerical findings and the conclusions can be found in Section \ref{ss-results}. In Subsections \ref{ss-averages} and \ref{ss-compare}, we study the averaged data and find further numerical validation of our method.
In Section \ref{s-supp} we verify the performance of the method in many examples. In Subsection \ref{ss-thouless}, we present the distribution of energies after repeated application of the random operator of a wave packet initially located at the origin.
We briefly remark on computing and memory requirements in Section \ref{s-final}.

\section{Preliminaries}\label{s-PRE}

\subsection{Singular and absolutely continuous parts of normal operators}\label{pre-normal}
Recall that an operator in a separable Hilbert space is called normal if $T^*T= TT^*$. By the spectral theorem operator $T$ is unitarily equivalent to $M_z$, multiplication by the independent variable $z$, in a direct sum of  Hilbert spaces
$$\cH = \oplus \int \cH(z)\, d\mu(z)$$ where $\mu$ is a scalar positive measure  on $\C$, called a scalar spectral measure
of $T$.

If $T$ is a unitary or self-adjoint operator, its spectral measure $\mu$ is supported on the unit circle or on the real line, respectively.
Via Radon decomposition, $\mu$ can be decomposed into a singular and absolutely continuous parts $\mu=\mu\ti{s}+\mu\ti{ac}$.
The singular component $\mu\ti{s}$ can be further split into singular continuous and pure point parts.
For unitary or self-adjoint $T$ we denote by $T\ti{ac}$  the restriction of $T$ to its absolutely continuous part, i.e.~$T\ti{ac}$ is unitarily equivalent to $M_t\big|_{\oplus \int \cH(t) d\mu\ti{ac}(t)}.$ Similarly, define the singular, singular continuous and the pure point parts of  $T$, denoted by $T\ti{s}$, $T\ti{sc}$ and $T\ti{pp}$, respectively.

\subsection{Key tool}
As mentioned above delocalization is deterministic. Therefore demonstrating that it does not occur with probability zero is sufficient to determine delocalization.

This following result makes our numerical experiment possible as it suffices to check the evolution of only one vector through repeated operations by the Anderson Hamiltonian and 3 dimensional random Schr\"odinger operator.

Fix the vectors $\delta_{000}\in l^2(\Z^3)$ and $\delta_{111}\in l^2(\Z^3)$, i.e.~3--tensors with zero entries, except for the $(0,0,0)-$position and the $(1,1,1)-$position, respectively, which equal $1$.

Notice that
\begin{align}\label{REF}
D_{\omega,c}^n := \dist(\delta_{111}, \text{span}\{H_\omega^k \delta_{000}:k=0,1,2,\hdots,n\})
\end{align}
describes the distance between the unit vector $\delta_{111}$ and the subspace obtained taking the closure of the span of the vectors $\delta_{000}, H_\omega \delta_{000}, H_\omega^2 \delta_{000}, \hdots, H_\omega^n \delta_{000}$.

In numerical linear algebra, this space is called a Krylov subspace,
and the Lanczos algorithm~\cite{lanczos1950iteration} provides a
classical approach for finding an orthonormal basis.  Our distance
calculation~\eqref{REF} relies on the orthogonality of these vectors,
iteratively updating the distance with each new Krylov vector.

\begin{prop}\label{c-tool}
Consider the discrete random Schr\"odinger operator given by equation \eqref{d-RandSchr}. Let $\omega_i$, $i\in \Z^3$, be i.i.d.~random variables with uniform (Lebesgue) distribution on $[-c,c]$, $c>0$. To prove delocalization (i.e.~the existence of absolutely continuous spectrum with positive probability), it suffices to find $c>0$ for which the distance
\begin{align}\label{e-dist}
D_{\omega,c}:=\lim_{n\to\infty}D_{\omega,c}^n =0
\end{align}
with non-zero probability. (Notice that the limit exists by the monotone convergence theorem.)
\end{prop}

The proposition follows immediately from Theorems 1.1 and 1.2 of \cite{JakLast2006} and is stated in more generality in \cite{2D}.

\begin{rem}\label{REMARK} The converse of Proposition \ref{c-tool} is not true. And we cannot draw any conclusions, if the distance between a fixed (unit) vector and the subspace generated by the orbit of another vector tends to zero. In particular, we cannot conclude that there must be localization. Even if we show \eqref{e-dist} for many or `all' vectors (instead of just $\delta_{111}$), it could be possible that the absolutely continuous part has multiplicity one and that $\delta_{000}$ is cyclic, that is, $l^2(\Z^3) = \clos\spa\{H_\omega^k \delta_{000}: k \in \N\cup\{0\}\}$.
\end{rem}

\section{Method of numerical experiment}\label{s-setup}
Consider the discrete Schr\"odinger operator given by \eqref{d-RandSchr} with random variable $\omega$ distributed according to the hypotheses of Proposition \ref{c-tool}.

By Proposition \ref{c-tool}, we obtain delocalization if we can find $c>0$ for which \eqref{e-dist} happens with non-zero probability. Let us now explain precisely how we verify delocalization numerically, leading up to the Numerical Criterion \ref{nc} below.

In the numerical experiment, we initially fix $c$ and fix one computer-generated realization of the random variable $\omega$ (with distribution in accordance to the hypotheses of Proposition \ref{c-tool}). We then calculate the distances $D_{\omega, c}^n$ for $n\in\{0, 1, 2, \hdots\}$.

Assuming that we know $D_{\omega,c}^n$ for $n=0, \hdots , 500$, let us find a lower estimate for the limit
$$
 D_{\omega,c} = \lim_{n\to\infty} D_{\omega,c}^n.
$$

\begin{figure}
\includegraphics[scale=0.6]{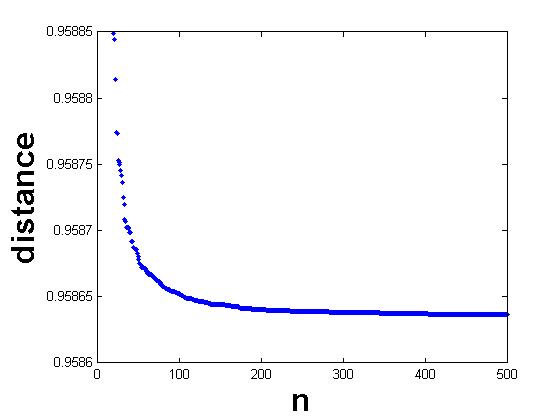}
\caption{Typical trend for $D^n_{\omega,c}$ for $n=500$ iterations and $c=0.3$.}
\label{c0}
\end{figure}

Figure \ref{c0} displays a typical trend for the distance $D_{\omega,c}^n$ as $n$ increases. Because the first $n=119$ points were irregular and do not contribute to the above limit, they were omitted. Notice that the graph is decreasing, as is expected. Although it certainly appears that the limit does not go to 0, the graph could have logarithmic decay, approaching zero very slowly. To attain an estimate for $D_{\omega,c}$, which excludes the case of such slow decay, we re-scaled the graph by $n ^{-a}$, $0.1 \leq a \leq 2$, so that the x-axis is inverted and the $y-$intercept, $y_{\omega,c}$, of a line of best fit will estimate $D_{\omega,c}$.

\begin{figure}
\includegraphics[scale=0.6]{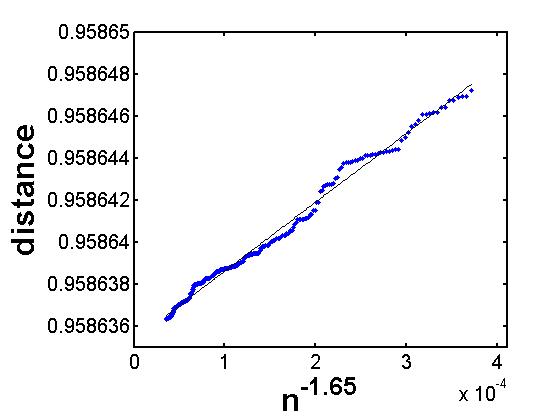}
\caption{Figure \ref{c0} re-scaled using $a=1.65$. Notice the fine $y-$scale and proximity to the $y-$axis.}
\label{c0a}
\end{figure}

Figure \ref{c0a} shows the re-scaled graph for $n = 120, 121, 122, \hdots, 500$. Subsection \ref{ss-a} describes the choice of $a$ and why, for small values of $c$, $D_{\omega,c}$ does not decay to 0.

\begin{figure}
\includegraphics[scale=0.6]{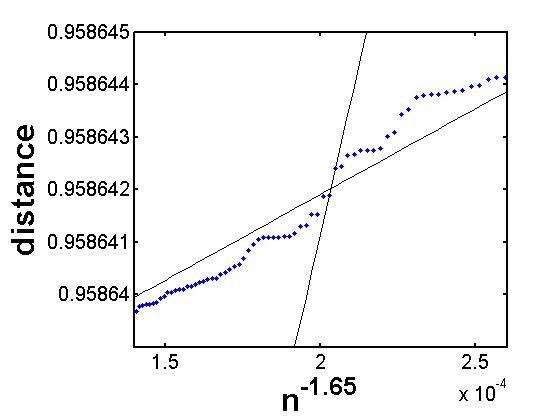}
\caption{A closer look at a window in Figure \ref{c0a}. The steeper line is used to determine $L_{\omega, c}$, and the line of best fit provides $y_{\omega, c}$. For this realization we have $L_{\omega,c}=0.9585894$ and $ y_{\omega,c} =  0.9586354$.}
\label{betterc0}
\end{figure}

Since an approximating line is only an estimate, for further confidence in our results, we also calculated the minimum $y-$intercept of all lines through two consecutive points and call it $L_{\omega,c}$ (see the steep line in Figure \ref{betterc0}). This is essentially the ``worst case," and ought to underestimate $D_{\omega,c}$, yielding the relationship
\begin{align*}
L_{\omega,c} \leq y_{\omega,c} \approx D_{\omega,c}\,.
\end{align*}

We repeated this process for many values of $c$ and multiple, different, computer-generated instances of the random variable $\omega$.
We took the minimum of $y_{\omega,c}$ and $L_{\omega,c}$ across all instances of $\omega$, with the intent to demonstrate that $D_{\omega,c}$ is above 0 for many different $\omega$.

In order to give confidence to our calculations to account for random error occurring in the computer, we introduce the following restrictions even though Proposition \ref{c-tool} only requires that $L_{\omega, c}>0$ .

\begin{criterion}\label{nc}
For a fixed value of $c$, we say that we have delocalization, if for at least 90\% realizations we obtain $L_{\omega, c}>.9>0$ and $y_{\omega, c}-L_{\omega, c}$ is of order $10^{-3}$. (Notice that we only need non-zero probability by Proposition \ref{c-tool}, and Remark \ref{REMARK}.)
\end{criterion}

\subsection{Choice of the re-scaling parameter}\label{ss-a}
For each fixed $c$ and $\omega$, the re-scaling exponent $a$ is chosen so that the re-scaled graph of the distance function (see Figure \ref{c0a}) satisfies the least square property; that is, the error with respect to square--norm when approximating the graph by a line is minimal. With this exponent we then find the corresponding linear approximation for the re-scaled distance function.

To find optimal $a$, we used the mesh $a=0.05:0.05:2$. Below is a table, see equation \eqref{table1}, for many values of $c$, giving the percentage of usable trials (those for which an optimal $a \geq 0.1$ was found) for each value of $c$.  Trials are not usable if the re-scaling parameter $a=0.05$ yields a concave graph. If this happens, we do not obtain any information (according to Remark \ref{REMARK}). See Figure \ref{c6} below. Note that a small value ($\leq 0.05$) of $a$ is ``bad", since the graph rescaled with $a=0.05$ will be concave, and thus it is not expected for a line of best fit to underestimate the limit of the distance.

\begin{figure}
\includegraphics[scale=0.6]{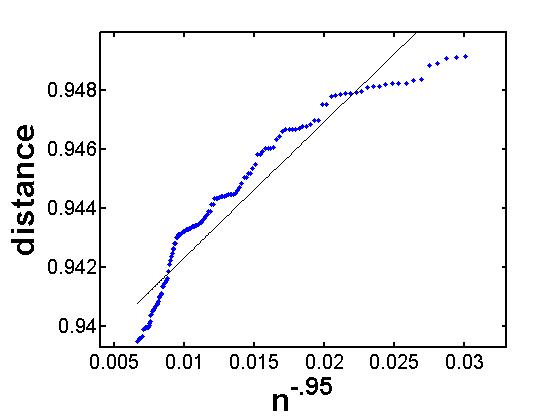}
\caption{A concave graph that yields no usable data ($c=6$). The concave shape of the data implies that $y_{\omega,c}$ is not necessarily a lower bound for $D_{\omega,c}$.}
\label{c6}
\end{figure}

A positive re-scaling factor implies that the graph in Figure \ref{c0} will not decay to zero. Indeed, using a re-scaling factor smaller than the optimal one will result in a convex graph for the distances $D_{\omega,c}^n$. And the $y-$intercept of the line lies below the value expected for $D_{\omega,c}^\infty$.

\section{Conclusions}\label{ss-results}
As mentioned in Section \ref{s-setup}, for a fixed $c$ we chose many realizations $\omega$. For every value of $c$, we took the minimum of the resulting quantities for $L_{\omega,c}$ and $y_{\omega,c}$ (the $y-$intercept of the approximating line and the minimum $y-$intercept of the lines passing through any two consecutive points, respectively).

We present our observations for the Numerical Criterion \ref{nc} for $n=500$. For fixed disorder, we will comment in Subsection \ref{ss-averages} on the re-scaling parameters of averages over the distances $D_{\omega, c}^n$, $n=0,1,2,\hdots, 200$ and $n=0,1,2,\hdots, 500$.

The following tables \eqref{table1} document the data obtained for n=200 by taking 15 realizations for each $c$ between $0.1$ and $5$, and 4 realizations for each $c\geq 10$ and $c=0$. By $\p$ we denote the probability of finding a re-scaling factor $a\in [0.1, 2]$.

\begin{align}
\label{table1}
\begin{array}{|c||c|c|c|c|c|c|c|}
\hline
c&0&0.1&0.2&0.3&0.4&0.5&0.6\\ \hline
\p&1&1&1&1&1&1&.93 \\ \hline
y_{\omega,c}&
.95869&
.95866&
.95859&
.95842&
.9582&
.9580&
.9573\\ \hline
L_{\omega,c}&
.95869&
.95865&
.95858&
.95738&
.9580&
.9563&
.9504\\ \hline
\end{array}
\end{align}

\begin{align*}
\begin{array}{|c||c|c|c|c|c|c|c|}
\hline
c&0.7&0.8&0.9&1&1.5&2&2.5\\ \hline
\p&.93&1&.93&.8&1&.87&1 \\ \hline
y_{\omega,c}&
.9573&
.9576&
.9559&
.9554&
.9540&
.9404&
.9447\\ \hline
L_{\omega,c}&
.9529&
.9574&
.9488&
.9487&
.9519&
.9063&
.9359\\ \hline
\end{array}
\end{align*}

\begin{align*}
\begin{array}{|c||c|c|c|c|c|c|c|}
\hline
c&3&3.5&4&4.5&5&10&15\\ \hline
\p&.8&.67&.73&.83&.6&1&.75 \\ \hline
y_{\omega,c}&
.9298&
.9105&
.8857&
.8843&
.8020&
.6458&
.0935\\ \hline
L_{\omega,c}&
.8443&
.7364&
.5389&
.5699&
.1095&
.3081&
-2.4974\\ \hline
\end{array}
\end{align*}

\begin{align*}
\begin{array}{|c||c|c|c|c|c|}
\hline
c&20&25&30&35&40\\ \hline
\p&.25&.75&.25&.75&.25 \\ \hline
y_{\omega,c}&
-2.1507&
-1.9412&
-3.3271&
-3.3301&
-9.7171\\ \hline
L_{\omega,c}&
-14.1608&
-11.8895&
-15.0577&
-38.6007&
-40.4666\\ \hline
\end{array}
\end{align*}

While for some $c\ge 2.5$, we have $\p\ge .9$ the difference between $y_{\omega,c}$ and $L_{\omega, c}$ is relatively large, which means that the line from taking the least square approximation is likely not a good approximation for the distances.

We also repeated the experiment for $n=500$ and the tables in equation \eqref{table2} below documents the findings. In these trials, the first 119 entries were removed instead of the first 44, as in the $n=200$ case. This larger crop makes the data more stable by giving better estimates for $y_{\omega,c}$ and $L_{\omega, c}$ and by more consistently finding a usable rescaling factor $a$. We ran 13 trials for $c\leq1$ and 4 trials for all other values.

\begin{align}
\label{table2}
\begin{array}{|c||c|c|c|c|c|c|c|}
\hline
c&0&0.1&0.2&0.3&0.4&0.5&0.6\\ \hline
\p&1&1&1&1&1&1&1 \\ \hline
y_{\omega,c}&
.95869&
.95866&
.95856&
.95838&
.95813&
.95782&
.9575\\ \hline
L_{\omega,c}&
.95869&
.95866&
.95855&
.95835&
.95809&
.95776&
.9574\\ \hline
\end{array}
\end{align}

\begin{align*}
\begin{array}{|c||c|c|c|c|c|c|c|}
\hline
c&0.7&0.8&0.9&1&1.5&2&2.5\\ \hline
\p&1&1&1&1&1&1&1 \\ \hline
y_{\omega,c}&
.9570&
.9565&
.9559&
.95523&
.9520&
.9518&
.9523\\ \hline
L_{\omega,c}&
.9569&
.9564&
.9558&
.95515&
.9514&
.9504&
.9438\\ \hline
\end{array}
\end{align*}

\begin{align*}
\begin{array}{|c||c|c|c|c|c|c|c|}
\hline
c&3&3.5&4&4.5&5&10&15\\ \hline
\p&1&1&1&1&1&.5&.75 \\ \hline
y_{\omega,c}&
.9451&
.9556&
.9405&
.9271&
.9149&
.2600&
-.5149\\ \hline
L_{\omega,c}&
.9417&
.9492&
.9302&
.9041&
.8055&
-2.3217&
-9.8398\\ \hline
\end{array}
\end{align*}

\begin{align*}
\begin{array}{|c||c|c|c|c|c|}
\hline
c&20&25&30&35&40\\ \hline
\p&.5&.5&.5&.75&.75 \\ \hline
y_{\omega,c}&
-2.8571&
-1.5957&
-2.5188&
-2.2407&
-3.1759\\ \hline
L_{\omega,c}&
-16.6379&
-24.0707&
-18.1168&
-18.7829&
-26.3506\\ \hline
\end{array}
\end{align*}

A good rescaling factor $a$ was found for all 143 of the trials for $c\leq 1$ and all $c\leq 3.5$ satisfy Criterion \ref{nc}, an improvement from the $n=200$ case. Hence the final conclusion of this numerical experiment is precisely the Main Result \ref{t-mr}.  According to Remark \ref{REMARK} and Criterion \ref{nc}, for $c\ge 4$, we do not have any conclusion.

\subsection{Averages}\label{ss-averages}
In the tables in equation \eqref{table3} below, for each fixed $c$, we averaged the distances $D_{\omega,c}^n$, $n=0, 1, 2, \hdots, 200$, of all our realizations. For those averaged distances, we determined the re-scaling parameters $\tilde a$, as well as $\tilde y_c$ and $\tilde L_c$ in analogy. The significance of our findings is that the re-scaling factors $\tilde{a}$ are ``roughly" decreasing and rather well-behaved for $c\le 1.5$. For larger disorder, $\tilde{a}$ becomes even less stable, and can't even be found for large enough disorder.

\begin{align}\label{table3}
\begin{array}{|c||c|c|c|c|c|c|c|}
\hline
c&0&0.1&0.2&0.3&0.4&0.5&0.6\\ \hline
\tilde{a}&2&1.9&1.65&1.5&1.3&1.1&.95 \\ \hline
\tilde{y}_{c}&
.95869&
.95869&
.95865&
.95861&
.95853&
.95846&
.95843\\ \hline
\tilde{L}_{c}&
.95869&
.95868&
.95864&
.95861&
.95852&
.95846&
.95841
\\ \hline
\end{array}
\end{align}

\begin{align*}
\begin{array}{|c||c|c|c|c|c|c|c|}
\hline
c&0.7&0.8&0.9&1&1.5&2&2.5\\ \hline
\tilde{a}&1.3&1&.8&.9&1&.6&.85 \\ \hline
\tilde{y}_{c}&
.95818&
.9584&
.9579&
.95779&
.95714&
.9539&
.9544\\ \hline
\tilde{L}_{c}&
.95816&
.9583&
.9578&
.95777&
.95708&
.9537&
.9541\\ \hline
\end{array}
\end{align*}

\begin{align*}
\begin{array}{|c||c|c|c|c|c|c|c|}
\hline
c&3&3.5&4&4.5&5&10&15\\ \hline
\tilde{a}&.65&.3&.55&.65&.5&.5&.3 \\ \hline
\tilde{y}_{c}&
.9485&
.9466&
.9345&
.9414&
.9217&
.8332&
.5312\\ \hline
\tilde{L}_{c}&
.9478&
.9442&
.9332&
.9399&
.9137&
.7648&
.2063\\ \hline
\end{array}
\end{align*}

\begin{align*}
\begin{array}{|c||c|c|c|c|c|}
\hline
c&20&25&30&35&40\\ \hline
\tilde{a}&.1&.85&.45&N/A&N/A \\ \hline
\tilde{y}_{c}&
-.3300&
.2928&
-.0722&
-3.0990&
-5.1751\\ \hline
\tilde{L}_{c}&
-1.8583&
-.0051&
-.4060&
-12.2712&
-12.7084\\ \hline
\end{array}
\end{align*}

In equation \eqref{table4} below we document the analogous quantities for the $n=500$ trials. Note that there is no rescaling factor for $c=20$, while there is for that $c$ in the $n=200$ trials. The data sets are not related to each other, aside from sharing the same disorder $c$.

\begin{align}
\label{table4}
\begin{array}{|c||c|c|c|c|c|c|c|}
\hline
c&0&0.1&0.2&0.3&0.4&0.5&0.6\\ \hline
\tilde{a}&2&1.75&1.35&1.2&1.1&1&1.05 \\ \hline
\tilde{y}_{c}&
.95869&
.95868&
.95864&
.95861&
.95855&
.95847&
.95823\\ \hline
\tilde{L}_{c}&
.95869&
.95868&
.95864&
.95861&
.95854&
.95846&
.95822\\ \hline
\end{array}
\end{align}

\begin{align*}
\begin{array}{|c||c|c|c|c|c|c|c|}
\hline
c&0.7&0.8&0.9&1&1.5&2&2.5\\ \hline
\tilde{a}&1.05&1.45&1.15&.65&.7&.6&.9 \\ \hline
\tilde{y}_{c}&
.95806&
.95786&
.95805&
.95795&
.9558&
.9543&
.9561\\ \hline
\tilde{L}_{c}&
.95805&
.95785&
.95803&
.95792&
.9555&
.9538&.
9556\\ \hline
\end{array}
\end{align*}

\begin{align*}
\begin{array}{|c||c|c|c|c|c|c|c|}
\hline
c&3&3.5&4&4.5&5&10&15\\ \hline
\tilde{a}&.55&.6&.65&.35&.15&.25&.3 \\ \hline
\tilde{y}_{c}&
.9506&
.9571&
.9479&
.9390&
.9244&
.7053&
.4991\\ \hline
\tilde{L}_{c}&
.9497&
.9531&
.9451&
.9348&
.9140&
.4796&
-.0327\\ \hline
\end{array}
\end{align*}

\begin{align*}
\begin{array}{|c||c|c|c|c|c|}
\hline
c&20&25&30&35&40\\ \hline
\tilde{a}&N/A&.1&.8&N/A&N/A \\ \hline
\tilde{y}_{c}&
.0569&
-1.9658&
-1.1264&
.0906&
-2.1984\\ \hline
\tilde{L}_{c}&
-.8254&
-6.5314&
-3.3376&
-.1823&
-6.8235\\ \hline
\end{array}
\end{align*}

\subsection{Comparing $n=200$ with $n=500$.}\label{ss-compare}
The $n=500$ data gave better results than the $n=200$ data. The probability of finding a useable rescaling factor for $n=500$ was higher than that of $n=200$ for all but two values of $c$. The average rescaling factor $\tilde{a}$ was similar between the two data sets. Finally, $y_{\omega,c}-L_{\omega,c}$ was smaller for the $n=500$ data for small $c$, suggesting that the approximation given by $y_{\omega,c}$ is better.

\section{Further validation of the method and the numerical experiments}\label{s-supp}
Apart from the usual tests (the program is running stably, checking all subroutines, many verifications for small $n$), we have conducted the following tests. Most important is the a posteriori test of orthogonality in the Lanczos algorithm in subsection \ref{ss-ortho}.

\subsection{Free discrete three dimensional Schr\"odinger operator}
When we apply the free discrete Schr\"odinger operator $H = H_{\bf 0}$ to the vector $\delta_{000}$, it immediately becomes clear that $H \delta_{000}$ as well as all vectors $H^n\delta_{000}$, $n\in\N\cup\{0\}$, are symmetric with respect to the origin. In dimension $d=3$, it is not hard to see that the distance between $\delta_{111}$ and the orbit of $\delta_{000}$ under $H$ is at least $ \frac{\sqrt{7}}{{2\sqrt{2}}}\approx 0.9354$. Indeed, we have
$$
\dist(\delta_{111}, \clos\spa\{H^n \delta_{000} :n\in\N\cup\{0\}\}) >
\min_x\dist(u_x, \delta_{111}) = \frac{\sqrt{7}}{{2\sqrt{2}}},
$$
where
$$
u_x = x\delta_{-1-1-1}+x\delta_{1-1-1}+x\delta_{-11-1}+x\delta_{-1-11}x\delta_{-1-11}+x\delta_{-11-1}+x\delta_{1-1-1}+x\delta_{111},
$$
the eight vertices of the length 2 cube centered at $(0,0,0)$.

In the experiments for the free discrete two dimensional Schr\"odinger operator we obtained a $y-$intercept of the approximating line approximately equals $0.9586936$. The re-scaled graph of distances still had a convex shape, so the actual distance as $n\to\infty$ would be bigger. In fact, we have extracted our data an upper estimate of $0.9586939 \approx D^{500}_{\omega, 0}$. Therefore, the distance must lie in the interval $[0.9586936, 0.9586939]$.

\subsection{Orthogonalization Process}

The $c=0$ case shows a decrease in distance on only every other step. The symmetry caused by the absence of random perturbations means the 3-tensor after orthogonalization has alternating diamonds of zero and nonzero entries radiating from the origin, meaning the distance decreases every second application of the operator, when there is a nonzero entry in the $(1,1,1)-$position.

\subsection{Evolution under $H_\omega$ of the bulk for small values of $c$}\label{ss-thouless}
We observe the bulk distribution which determines the distance from the origin where we are most likely to find an electron. Here, distance is measured by the taxicab method, so elements of the same distance form a diamond in the 3-D integer lattice. The bulk at this distance is the Euclidean norm of the elements constituting the diamond.

To be precise, we consider the elements of the vector $m_{500}$ and define
\begin{align}\label{e-E}
E(l,n) = \sqrt{\sum_{|i|+|j|+|k|=l} (m_n)_{i,j,k}^2}
\end{align}
for the bulk $E(l,n)$ of the vector $m_n$ at taxicab distance $l$ from the origin. Here $(m_n)_{i,j,k}$ refers to the $(i,j,k)-$entry of the $2-$tensor $m_n$. Slightly abusing notation, we normalize $m_n$ and use the same notation for the normalized sequence of vectors.

\begin{figure}
\includegraphics[scale=0.45]{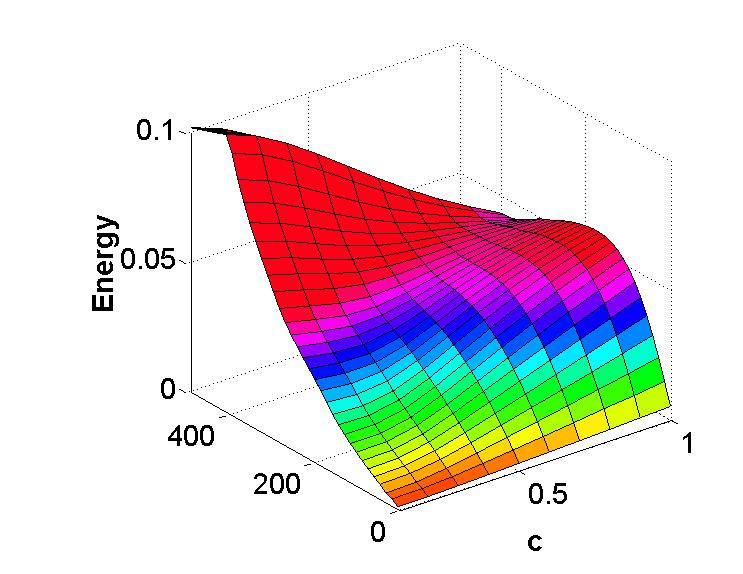}
\caption{Energy distribution of $m_{500}$ for the disorders $c=0.1 : 0.1 :1$, averaged over the 11 realizations for each value of $c$.}
\label{n500energya}
\end{figure}

Figure \ref{n500energya} is the result of averaging four sets of data for values of $c$ ranging from $0.1$ to $1$. As expected, the energy remains closer to the origin as disorder increases.

\subsection{Lanczos and orthogonality}\label{ss-ortho}
The Lanczos algorithm is known to lose orthogonality in many instances,
which could cast doubt on our distance calculations.  To test the
accuracy for our problem, we stored the entire Krylov subspace
generated on a smaller problem instance ($N=150$) and stored these as
columns of a matrix $K$.  The quantity $Q = \left\| K^T K - I \right\|_\infty$
should deviate with zero in proportion to the loss of orthogonality.
In Tables~\eqref{Q1} and \eqref{Q2}, we measure the matrix $\infty$ norm for realizations
for several cases of $c$ near the critical value.  We see that the
Krylov vectors in these cases are in fact quite close to orthogonal especially for $c\le3.5$,
although the orthogonality seems to decrease as $c$ grows.

\begin{align}\label{Q1}
\begin{array}{|c|c|c|c|c|c|c|c|c|c|c|c|c|c|c|c|c|c|}
\hline
c & 
0.0 & 
0.5 & 
1.0 & 
1.5 &
2.0 & 
2.5 & 
3.0 & 
3.5 \\
\hline
Q&
2.01e{-11} &
7.2e{-11}&
4.8e{-11}&
 4.4e{-11}&
 6.9e{-11} &
 9.6e{-11} &
 3.1e{-11}&
  4.8e{-11}\\\hline
\end{array}
\end{align}

\begin{align}\label{Q2}
\begin{array}{|c|c|c|c|c|c|c|c|c|c|c|c|c|c|c|c|c|c|}
\hline
c & 
4.0 & 
4.5 & 
5.0 &
5.5 & 
6.0 & 
6.5 & 
7.0 & 
7.5 & 
8.0 \\
\hline
Q&
  4.5e{-11} &
  2.4e{-11} &
   6.1e{-9} &
   1.3e{-10} &
   4.0e{-10} &
   9.2e{-11} &
   1.2e{-9} &
   7.3e{-8} &
   5.1e{-9} \\\hline
\end{array}
\end{align}

\section{On computing and memory requirements}\label{s-final}

Using methodology similar to that in \cite{2D}, all of the information contained in the 3-tensor is stored in one information vector. For this method, because of how the Hamiltonian acts, it is important for computing purposes that each point in the 3-tensor is stored in a position such that its neighbors along a coordinate axis are a consistent distance from that point in the vector. This methodology allows the vector to be half the size necessary for containing every point in a 3-tensor, but still approximately twice as large as is necessary. In order to explore localization in higher dimensions, a more efficient method is needed since a generalization of this code for dimension $d$ has time complexity $\mathcal{O}(n^d)$.

After prototyping our approach in MATLAB, we translated the code into
FORTRAN90.  This
allowed us a smaller memory footprint and hence larger and more
efficient runs.  We then wrapped this routine into Python using the
\texttt{f2py} package~\cite{peterson2009f2py}.  By doing so, we were
able to run several cases concurrently on our workstation by using
Python's \texttt{multiprocessing} module.

Our simulations were run on a Dell Precision workstation with dual
eight-core Intel Xeon E5-2680 processors running at 2.7GHz with 128GB of
RAM.  We used gfortran version 4.4.7 with flags
\texttt{-O3 -ftree-vectorizer-verbose=2 -msse2 -funroll-loops
-ffast-math},\\which, among other optimizations, enables
instruction-level superscalar parallelism.

\section{Further Projects}
An immediate area for further exploration would be to consider various geometries, rather than simply the n-dimensional lattice. One geometry of interest is the Sierpinski gasket, starting at one corner and building the various triangles as $n$ increases. Preliminary results indicate that a program modeling the free random Schr\"odinger operator on this geometry should run with time complexity $\mathcal{O}\left(n^{\frac{\ln(3)}{\ln(2)}}\right).$

\providecommand{\bysame}{\leavevmode\hbox to3em{\hrulefill}\thinspace}
\providecommand{\MR}{\relax\ifhmode\unskip\space\fi MR }
\providecommand{\MRhref}[2]{%
  \href{http://www.ams.org/mathscinet-getitem?mr=#1}{#2}
}
\providecommand{\href}[2]{#2}

\end{document}